\documentclass[aps,prd,twocolumn,showpacs,superscriptaddress]{revtex4-1}

\usepackage{graphicx}  
\usepackage{bm}        
\usepackage{amssymb}   
\usepackage{amsmath}   
\usepackage{multirow}
\usepackage{subcaption}

\usepackage{xcolor}

\newcommand*{\snoplus}{SNO\raisebox{0.35ex}{\small\textbf{+}}}
\graphicspath{{./figures/}}

\begin{document}

\title{Improved search for invisible modes of nucleon decay in water with the \snoplus{} detector}

\author{\bf A.\,Allega}
\affiliation{\it Queen's University, Department of Physics, Engineering Physics \& Astronomy, Kingston, Ontario K7L 3N6, Canada}
\author{\bf M.\,R.\,Anderson}
\affiliation{\it Queen's University, Department of Physics, Engineering Physics \& Astronomy, Kingston, Ontario K7L 3N6, Canada}
\author{\bf S.\,Andringa}
\affiliation{\it Laborat\'{o}rio de Instrumenta\c{c}\~{a}o e  F\'{\i}sica Experimental de Part\'{\i}culas (LIP), Av. Prof. Gama Pinto, 2, 1649-003, Lisboa, Portugal}
\author{\bf M.\,Askins}
\affiliation{\it University of California, Berkeley, Department of Physics, California 94720, Berkeley, USA}
\affiliation{\it Lawrence Berkeley National Laboratory, 1 Cyclotron Road, Berkeley, California 94720-8153, USA}
\author{\bf D.\,J.\,Auty}
\affiliation{\it University of Alberta, Department of Physics, 4-181 CCIS,  Edmonton, Alberta T6G 2E1, Canada}

\author{\bf A.\,Bacon}
\affiliation{\it University of Pennsylvania, Department of Physics \& Astronomy, 209 South 33rd Street, Philadelphia, Pennsylvania 19104-6396, USA}
\author{\bf N.\,Barros}
\affiliation{\it Laborat\'{o}rio de Instrumenta\c{c}\~{a}o e  F\'{\i}sica Experimental de Part\'{\i}culas (LIP), Av. Prof. Gama Pinto, 2, 1649-003, Lisboa, Portugal}
\affiliation{\it Universidade de Lisboa, Faculdade de Ci\^{e}ncias (FCUL), Departamento de F\'{\i}sica, Campo Grande, Edif\'{\i}cio C8, 1749-016 Lisboa, Portugal}
\author{\bf F.\,Bar\~{a}o}
\affiliation{\it Laborat\'{o}rio de Instrumenta\c{c}\~{a}o e  F\'{\i}sica Experimental de Part\'{\i}culas (LIP), Av. Prof. Gama Pinto, 2, 1649-003, Lisboa, Portugal}
\affiliation{\it Universidade de Lisboa, Instituto Superior T\'{e}cnico (IST), Departamento de F\'{\i}sica, Av. Rovisco Pais, 1049-001 Lisboa, Portugal}
\author{\bf R.\,Bayes}
\affiliation{\it Laurentian University, Department of Physics, 935 Ramsey Lake Road, Sudbury, Ontario P3E 2C6, Canada}
\author{\bf E.\,W.\,Beier}
\affiliation{\it University of Pennsylvania, Department of Physics \& Astronomy, 209 South 33rd Street, Philadelphia, Pennsylvania 19104-6396, USA}
% ...
\collaboration{The SNO+ Collaboration}

\date{\today}
\begin{abstract}
This paper reports results from a search for single and multinucleon disappearance from the $^{16}$O nucleus
in water within the \snoplus{} detector using all of the available data. These so-called
``invisible'' decays do not directly deposit energy within the detector but are instead detected through
their subsequent nuclear deexcitation and gamma-ray emission. New limits are given for the partial lifetimes:
$\tau(n\rightarrow inv) > 9.0\times10^{29}$ years, $\tau(p\rightarrow inv) > 9.6\times10^{29}$ years,
$\tau(nn\rightarrow inv) > 1.5\times10^{28}$ years, $\tau(np\rightarrow inv) > 6.0\times10^{28}$ years, and
$\tau(pp\rightarrow inv) > 1.1\times10^{29}$ years at 90\% Bayesian credibility level (with a prior uniform in rate). 
All but the ($nn\rightarrow inv$) results improve on existing limits by a factor of about 3.
\end{abstract}

\maketitle
\section{Introduction}
\label{sec:Introduction}
Baryon number and lepton number are accidental symmetries of the Standard Model, yet to date no 
process violating baryon number and/or lepton number conservation has been observed. Any proven 
violation of the baryon number conservation principle would be an important step toward explaining 
the apparent matter-antimatter asymmetry of the Universe. The decay of nucleons would provide a 
direct observation of baryon number conservation violation and has been the goal of many experiments for the past several decades
\cite{KamiokandeNDecay,KamiokandeNDecay2,KamiokandeNDecay3}. The particulars of the possible
decay modes and their various branching ratios depend heavily on the chosen model, so we
seek to make a model-independent measurement of the partial lifetime through individual
decay modes. So-called ``invisible'' decay modes are those where the decay daughter particles cannot be directly observed,
but the nuclei are left in an excited state whose deexcitation does deposit energy into the detector. In 
water, this signal manifests from the energy deposited by gamma-rays in the 5--10 MeV energy window, to which
the \snoplus{} detector is particularly sensitive, owing to its low backgrounds. The decay modes of interest include
the disappearance of a single nucleon ($n$ or $p$) as well as the simultaneous disappearance of a pair
of nucleons ($nn$, $np$, or $pp$). Previous limits have been set by
SNO \cite{SNONucleonDecay}, KamLAND \cite{KamLANDNucleonDecay}, and \snoplus{} \cite{FirstNDPaper}, 
and this paper presents improved results on those previously published using additional data, reduced background, and improved analysis techniques.
\section{The SNO\raisebox{0.5ex}{\tiny\textbf{+}} Detector}
\label{sec:Detector}

The \snoplus{} detector is located underground in Sudbury, Ontario, at a depth of
2070 meters. The detector, which is currently filled with liquid scintillator, had about 1kT 
target volume of ultrapure water in its first phase of data-taking. This target volume was 
contained in a 6-meter radius spherical acrylic vessel (AV), which was
surrounded by an additional buffer volume of ultrapure water and 9394 8-inch photomultiplier
tubes (PMTs) mounted on a steel support structure 9-meters from the center of the AV.
The AV is held in suspension through a network of Tensylon ropes anchored both to the
laboratory deck above the AV and to the cavity floor. A 7-meter long, 1.6 meter diameter acrylic tube attached at
the top of the AV extends the active volume up to the laboratory deck, where the water comes
into contact with the gas above it.

The first set of published data from the \snoplus{} water phase spanned various
periods of detector and water system commissioning between May 2017 and December 2017, and
as a result included variable levels of radon-induced background \cite{FirstNDPaper}. 
The data presented here includes a subset of the previous data, as well as an additional set of low background data that was
taken after the installation of an AV cover gas system, between October 2018 and July 2019. 
Radon and its daughters can be drawn into the target volume both through diffusion and convection from the
liquid/gas interface at the top of the detector. The AV cover gas is a sealed system that uses 
nitrogen gas to protect the volume above the water in the detector against 
radon, reducing the radon ingress by at least an order of 
magnitude. Further details on the detector can be found in \cite{SnoplusDetector, SnoplusOptics}.
\section{Data Selection}
\label{sec:DataSelection}
The water phase of
the experiment represents a transitional period in which the detector was commissioned
and prepared to be filled with liquid scintillator. During this phase, the background
levels fluctuated due to changes in the water processing and recirculation system, and
the data was divided into periods of relatively stable background levels. 
The ingress of radon into the detector produces an inhomogeneous field of background 
activity within the detector.

In this analysis, the fiducial volumes in the different periods were reoptimized to further
suppress this variable background. 
Since almost all of the transient radon appeared near the top of the detector, this was done by
comparing the distribution of events in an energy side-band sensitive to radon (4--5 MeV) between
the upper and lower hemispheres to estimate the population of events related to transient radon
distributions in different fiducial volumes. The fiducial volume was then optimized by balancing 
the statistical uncertainty and the systematic uncertainty related to the residual transient radon 
events in the fiducial volume.
The previously published data consisted of six distinct sets of
data, each with their own selected fiducial volumes. In all cases, the new fiducial volume
is smaller than the previous one, and one of the datasets was removed entirely
for this analysis \cite{FirstNDPaper}. A breakdown of the individual dataset livetimes and their fiducial volumes
is given in Table \ref{tab:evselection}. The total livetime across all time periods is $274.7\pm1.0$
days.


\begin{table*}
\centering
\begin{tabular}{c||c|c|c|c|c|c}
\hline \hline
Observable & Period 1 & Period 2 & Period 3 & Period 4 & Period 5 & Period 6\\
\hline
$R$ (m) max & 5.1 & 5.1  & 5.1  & 5.1  & 5.4  & 5.2  \\
Z (m) [min, max]   & [-6.0, 1.5] & [-6.0, 1.5] & [-6.0, 1.5] & [-3.1, 1.9] & [-6.0, 2.0] & [-6.0, 3.0] \\
\hline
Livetime (days) & 5.0 & 14.6 & 30.2 & 28.9 & 11.2 & 184.8 \\
\hline \hline
\end{tabular}
\caption{Optimized fiducial volume and livetime for each of the included datasets.}
\label{tab:evselection}
\end{table*}
\section{Event Reconstruction}
\label{sec:Reconstruction}
Detailed optical calibrations of the detector were carried out, and an updated optical model
including a new tabulation of the optical attenuation lengths of the AV and water volumes,
was developed and applied to the new dataset, details of the calibration process and comparison
with the $^{16}$N calibration source can be found in \cite{SnoplusOptics}.
Additionally, the collective angular
response of the PMTs and light concentrators was updated following improved calibrations.
This new modeling was only applied to the newly acquired period 6 data and associated simulations, resulting in 
different reconstruction systematic uncertainties for the old and new data.

Event reconstruction does not differ greatly from the previous results, and involves reconstructing the most likely position ($\Vec{\mathrm{R}}$), direction ($\Vec{\mathrm{U}}$, orientation of the momentum vector), and effective electron-equivalent kinetic energy ($\mathrm{T}_e$) of an event in the detector based on the arrival time of photons at the PMTs, assuming those photons to be electron generated Cherenkov light.
Two additional quantities, the in-time-ratio (ITR) and the light isotropy $\beta_{14}$,
were used to help identify poor fits and nonphysical events (such as those caused by the
detector electronics). These two are as previously described in \cite{FirstNDPaper} and 
the same selection criteria of $\mathrm{ITR} > 0.55$ and $-0.12 < \beta_{14} < 0.95$ in \cite{FirstNDPaper} are retained
for this analysis.
Additionally, several new reconstruction figures of merit were used to identify poorly 
reconstructed event positions and energies.


Systematic uncertainties arising from the detector modeling and event reconstruction were evaluated
using $^{16}$N calibration data. Changes in the optical calibration between the
previous dataset (periods 1 through 5) and the newest dataset (period 6) require evaluation
of that calibration data using the two different models. The improved optical modelling used in the
newer dataset improved the energy scale and resolution uncertainty by a factor of 2.
Additionally, the uncertainty in the calibration source position was applied as an offset 
in the newer dataset as opposed to folding it in with the overall position scale uncertainty. 

A summary of all of the systematic uncertainties, split by data-taking period, is given in Table \ref{tab:systematics}.
\begin{table*}
    \centering
    \begin{tabular}{c|c|c}
        \hline \hline
        Parameter & Uncertainty (1 - 5) & Uncertainty (6)            \\  
        \hline
        x offset (mm) &  ${}^{+16.4}_{-18.2}$ & ${}^{+50.1}_{-55.6}$ \\
        y offset (mm) &  ${}^{+22.3}_{-19.2}$ & ${}^{+47.7}_{-59.6}$ \\
        z offset (mm) &  ${}^{+38.4}_{-16.7}$ & ${}^{+75.8}_{-34.7}$ \\
        \hline
        \multirow{2}{*}{x scale (\%)}  &  \multirow{2}{*}{${}^{+0.91}_{-1.01}$} & $(x>0)~~{}^{+0.16}_{-0.23}$ \\
                                       &                                        & $(x<0)~~{}^{+0.17}_{-0.30}$ \\
        \multirow{2}{*}{y scale (\%)}  &  \multirow{2}{*}{${}^{+0.92}_{-1.02}$} & $(y>0)~~{}^{+0.12}_{-0.22}$ \\
                                       &                                        & $(y<0)~~{}^{+0.17}_{-0.45}$ \\
        \multirow{2}{*}{z scale (\%)}  &  \multirow{2}{*}{${}^{+0.91}_{-0.99}$} & $(z>0)~~{}^{+0.30}_{-0.42}$ \\
                                       &                                        & $(z<0)~~{}^{+0.09}_{-0.24}$ \\
        \hline
        x resolution (mm) &  104 & $\sqrt{3214 + |0.393x - 290|}$ \\
        y resolution (mm) &  98  & $\sqrt{2004 + |0.809y - 1365|}$ \\
        z resolution (mm) &  106 & $\sqrt{7230 + |0.730z + 3211|}$ \\
        \hline
        Angular resolution & ${}^{+0.13}_{-0.08}$ & ${}^{+0.122}_{-0.020}$\\
        \hline
        $\beta_{14}$ & ${}^{+0.003}_{-0.010}$ & ${}^{+0.005}_{-0.010}$\\
        \hline
        Energy scale (\%) & 2.0 & 1.02 \\
        Energy resolution & ${}^{+0.018}_{-0.016}$ & ${}^{+0.0084}_{-0.0079}$\\
        \hline \hline
    \end{tabular}
    \caption{
    Summary of the evaluated systematic uncertainties for the reconstructed
    parameters, for the various datasets. Due to updates in the optical modeling,
    the final dataset (6) has a separate evaluation of these uncertainties from
    the previous datasets (1 - 5)\cite{FirstNDPaper}.
    }
    \label{tab:systematics}
\end{table*}

\begin{figure}
    \centering
    \includegraphics[width=\linewidth]{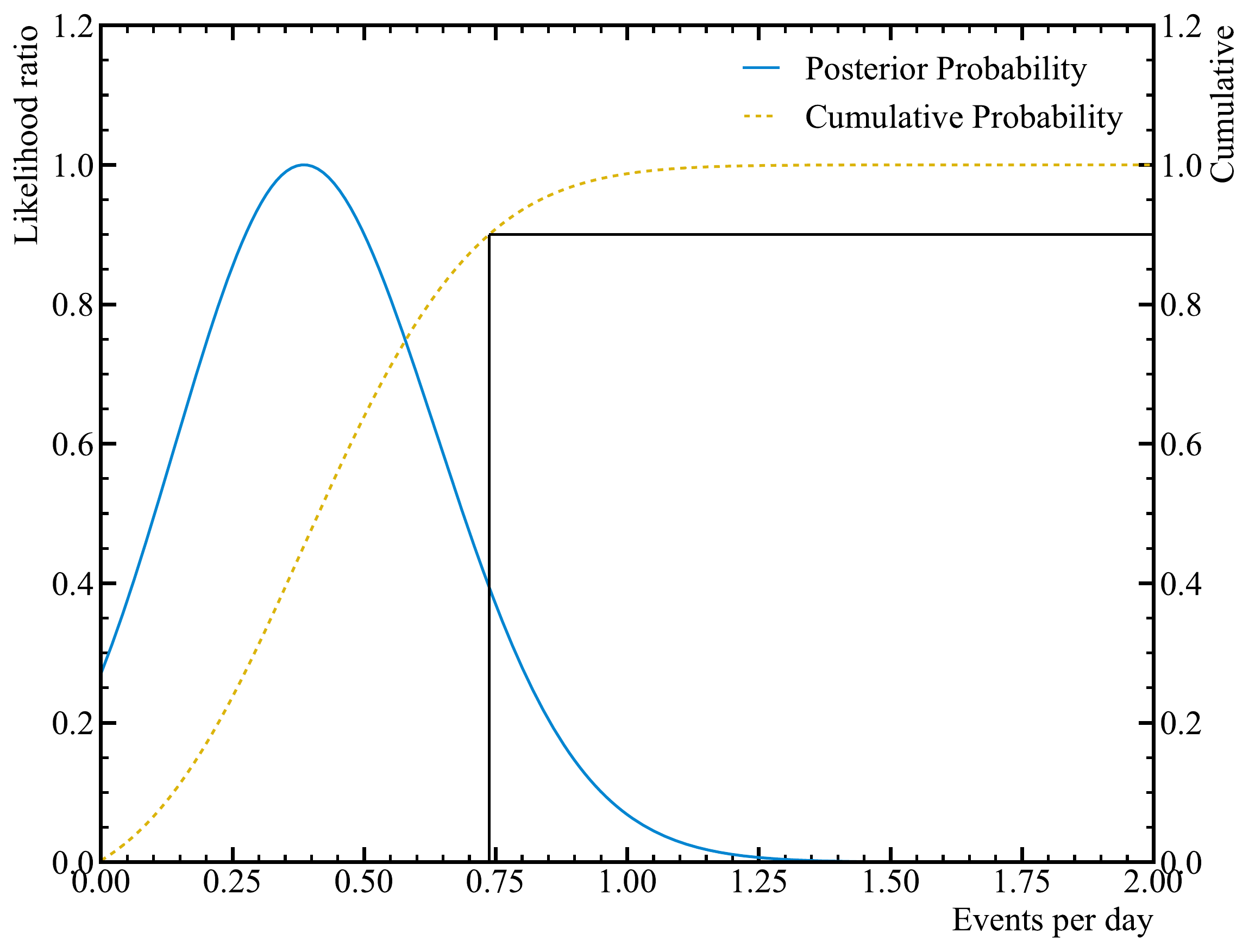}
    \caption{The likelihood ratio for the neutron decay mode versus the limit of the number
    of signal decays per day. The associated cumulative distribution indicates the point
    at which the limit on the signal at 90\% C.I. is drawn.}
    \label{fig:likelihood}
\end{figure}
\section{Background Model}
\label{sec:Backgrounds}
For each of the hypothetical decay modes considered here, the primary detector signature would be one or more
gammas depositing a characteristic quantity of energy. In order to identify such events, it is necessary 
to understand (and thereby account for) various sources of background within the chosen energy window (5--10 MeV).

\subsection{Instrumental background}
Low-level data checks were performed to identify and remove spurious events due to instrumentation
and other non-physics sources. These data-cleaning cuts entail a small signal sacrifice and
an associated contamination that must be taken into account. The signal sacrifice was determined using the
$^{16}$N calibration source for each dataset, and varies due to the different fiducial volumes.
Residual contamination was estimating using 
a data-driven analysis which compared two sets of data-cleaning cuts using a
bifurcated analysis technique. In this technique, the data was separated into 4 categories based
on whether it passed or failed two branches of data-cleaning. One branch
consisted of a set of low-level data-cleaning cuts based on the hit-times and charge recorded for an
event, while the other used the reconstruction classifiers ($\beta_{14}$ and ITR). Based on the
individual leakage of these two branches the resulting sacrifice and contamination for each of the
datasets is given in Table \ref{tab:datacleaning}. The estimated contamination and sacrifice with the
updated reconstruction and systematics are consistent with previous estimates using this technique \cite{FirstNDPaper}.

\begin{table}[!h]
    \centering
    \begin{tabular}{c|c|c}
        \hline \hline
        Data set & Sacrifice & Contamination (events) \\
        \hline
        1 & $(1.2 \pm 0.3)\%$ & $0.01 \pm 0.02$ \\
        2 & $(1.2 \pm 0.3)\%$ & $0.04 \pm 0.03$ \\
        3 & $(1.2 \pm 0.3)\%$ & $0.03 \pm 0.03$ \\
        4 & $(1.2 \pm 0.3)\%$ & $0.08 \pm 0.06$ \\
        5 & $(1.3 \pm 0.3)\%$ & $0.01 \pm 0.01$ \\
        6 & $(1.7 \pm 0.4)\%$ & $0.04 \pm 0.01$ \\
        \hline \hline
    \end{tabular}
    \caption{Fraction of signal sacrificed by data-cleaning cuts and the corresponding
    number of nonphysical events expected to pass the signal selection criteria,
    as evaluated from the physics data and calibration data.}
    \label{tab:datacleaning}
\end{table}

\subsection{Cosmogenic background}
The depth of the \snoplus{} detector reduces the total muon flux in the detector to only about 3 muons per hour. 
This is a vital advantage in the search for invisible nucleon decay, since muon spallation reactions producing $^{16}$N 
yield an event signature that is almost identical to the nucleon decay signal. 
Therefore, it is important to remove the remaining cosmogenic $^{16}$N events by muon identification and a 
20-second cut on events after each detected muon. Based on an estimated cosmogenic rate from \cite{SuperKCosmogenic, SuperKCosmogenic2},
scaled to the \snoplus{} detector depth, and the 20-second muon veto cut, there is 1 event expected within
the fiducial volume, across the cumulative livetime of the datasets.
\subsection{Radioactive background}
The dominant background in this analysis is due to trace levels of residual contamination from uranium and thorium, 
as well as radon ingress, that primarily manifest as $^{214}$Bi and $^{208}$Tl decays. The majority of these events
fall below the 5--MeV analysis threshold.

The radioactivity intrinsic to the internal water was fit as two individual components, $^{214}$Bi and $^{208}$Tl. The 
presence of the AV cover-gas system reduced the internal $^{214}$Bi (U-chain) and $^{208}$Tl (Th-chain) levels by one 
order of magnitude compared to the previous analysis \cite{FirstNDPaper}, with current values of 
(5.78$\pm$0.7$^{+1.5}_{-1.3}$)$\times10^{-15}$ gU/g and $<4.8\times10^{-16}$ gTh/g (95\% C.L.), respectively.

For this analysis, all of the decays external to the target volume were treated as a single parameter in the
fit. This single parameter includes contributions from decays in the external water, the AV,
and the ropes that hold the AV in place.
Glass in the PMTs has higher radioactivity relative to other detector components, but the approximately two and 
a half meters of external water between the PMTs and the fiducial volume attenuates this contribution to detector background 
to a negligible level, and so this background was not included in the fit.

\subsection{Neutrino background}
Several sources of neutrino and antineutrino background were present, which were constrained
through previous experimental results and data. The most significant source of background comes from $^{8}$B solar
neutrinos, which have been well measured. For these results, the solar neutrino flux is constrained
based on recent results from Super-Kamiokande \cite{SuperKSolar}, which are consistent with both the
SNO \cite{SNOSolar} and SNO+ \cite{SNOPlusSolar} measurements, with oscillations applied using
the BS2005-OP solar model \cite{BS2005-OP}. 

Atmospheric neutrino interactions also 
create signals within the detector through neutral current interactions that can liberate nucleons from
the oxygen nucleus. The subsequent deexcitation of $^{15}\mathrm{N}^*$ or $^{15}\mathrm{O}^*$ could look
identical to the single proton and neutron decay modes. However, many of these interactions can
be identified by detecting the neutron captures that follow the event. Estimates based on GENIE
simulations \cite{GENIE} predict about 0.1 interactions produce an excited state nucleus per day within the 
entire internal water volume. By including the uncertainty on the atmospheric neutrino flux,
GENIE branching ratio, and the interaction cross-section, a 67\% total uncertainty on the normalization 
of the atmospheric contamination rate was found.

Antineutrinos from the nearby nuclear reactors were also considered in the analysis, though
their overall contribution was relatively small. The flux estimates combined reactor information from the
International Atomic Energy Agency \cite{IAEA} with Canadian reactor power information \cite{IESO} to provide a constraint.
Furthermore, because the reactor signals are inverse beta decays, the neutron follower cut designed 
to reduce the number of atmospheric events served to also reduce the reactor antineutrino background.
The total number of reactor antineutrino background events was fit in the analysis with the included
constraint from the flux, yielding an estimate of 1.8 events summed across all of the datasets within the fiducial
volumes.

\subsection{Additional subdominant backgrounds}
Other backgrounds were studied and found to be negligible in magnitude and not included in the fit.
These backgrounds would not normally fall within the region of interest, but can occasionally do so due to
misreconstruction.
From Monte Carlo simulations, the PMT background is expected to contribute fewer than 0.1 events within
the region of interest in the entire dataset. The other considered sources of subdominant background were ($\alpha$,n)
reactions on $^{13}$C (acrylic) or $^{18}$O (acrylic and water) nuclei. In the $^{13}$C case, an $\alpha$ is 
captured on the nucleus, resulting in a neutron and a $^{16}$O nucleus, which can be produced in an excited 
state of 6.1\,MeV up to 10\% of the time. The gammas or electrons/positrons emitted in the deexcitation have a small
chance of falling into the region of interest (ROI) for this study. However, the reaction on $^{18}$O produces low energy 
gammas and, therefore, it is expected to make a negligible contribution in the ROI. The dominant source of alphas is from 
$^{210}$Po embedded a few nanometers below the AV surface \cite{SnoplusDetector}. 

In total, there are expected to be fewer than 1.4 events out of a total of 239 in the dataset that are not
included in the fit. The expected contribution from these backgrounds are not subtracted from
the final signal count, producing a more conservative upper limit on the nucleon decay lifetimes.

\begin{figure*}
\begin{subfigure}{0.45\textwidth}
    \includegraphics[width=\linewidth]{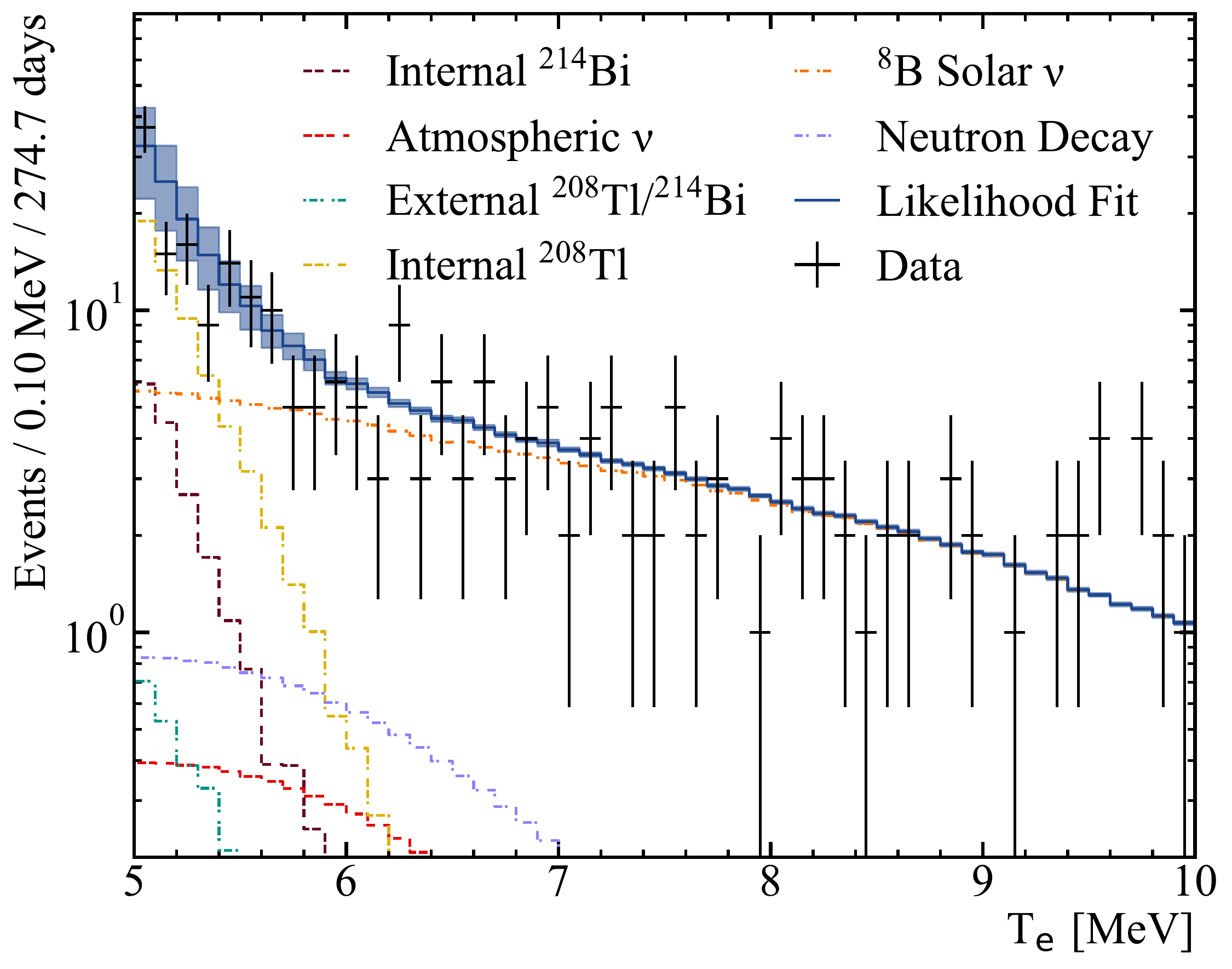}
\end{subfigure}
\begin{subfigure}{0.45\textwidth}
    \includegraphics[width=\linewidth]{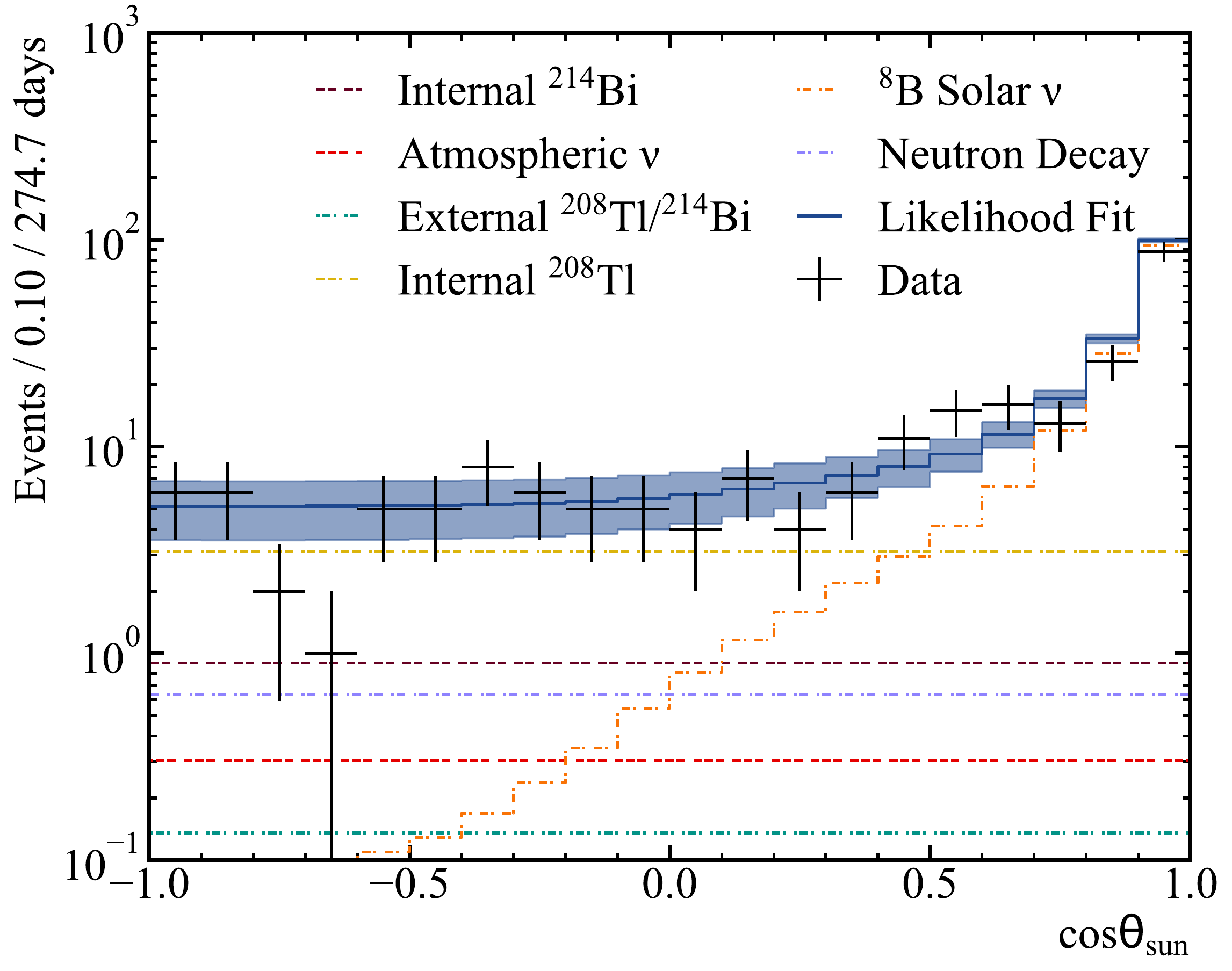}
\end{subfigure}
    \caption{Projection of the reconstructed energy (left) and the dot product
    of the reconstructed direction and the solar direction (right) for the entire
    dataset. The error bands on the total fit show the result of all of the
    systematic uncertainties applied. The neutron decay signal shown is at the best-fit
    lifetime of $1.7 \times 10^{30}$ years.
    }
    \label{fig:energyfit}
\end{figure*}
\section{Analysis Methods}
\label{sec:Analysis}
This analysis applied new selection criteria to the previous data (consisting of
datasets 1 through 5), with the addition of the sixth blinded dataset.
Data was blinded in the approximate energy range of 5 to 15 MeV based on the number of PMT hits
per event.
The region outside of the blinded region (unblinded data) was used for a fiducial volume selection study. 
As a result of radon ingress from the gas/liquid interface above, the spatial field of background 
radioactivity in the detector was not homogeneous, and so it was necessary to limit the fiducial volume by
placing a varying cut in Z for each of the datasets independently. Since these asymmetries are not modeled in
Monte Carlo simulations, an analysis on data outside of the region of interest, from 4--5 MeV, was used to assess
this top-bottom asymmetry and select a fiducial volume that would minimize
the uncertainty due to this event contamination. The event selection criteria for each
of the individual datasets is given in Table \ref{tab:evselection}. After applying the selection criteria and
each decay mode's individual branching ratio, the final signal acceptance efficiencies are given in Table \ref{tab:efficiency}.

Events within the selected ROI between 5 and 10 MeV were fitted using an unbinned
maximum likelihood method with probability distributions produced from the Monte Carlo
in energy, position, and direction. The normalization on the signal rate and the rate of internal and external radioactive
backgrounds were fit unconstrained, while the cosmogenic $^{16}\mathrm{N}$, atmospheric neutrino, and solar neutrino events
included a Gaussian constraint as discussed in the previous section. The fitting technique
was repeated for each of the signals of interest ($n$, $p$, $pp$, $np$, and $nn$ decay)
independently. The Monte Carlo model simulates events uniformly in
time throughout the data taking window, including variations in the detector state. 
The probability distribution functions (PDFs) were separated into a one-dimensional distribution
in $R^3$ multiplied by a two-dimensional distribution in $(\cos{\theta_\mathrm{sun}}, \mathrm{T}_e)$,
where $\cos{\theta_\mathrm{sun}}$ is the dot product between the reconstructed event
direction and the position of the sun with respect to the event position.
The choice to split the full PDF this way was due to the strong correlation between $\cos{\theta_\mathrm{sun}}$
and $\mathrm{T}_e$ found in the solar neutrino background (due to the decreasing opening angle
between the electron and the incident neutrino in the elastic scattering process), 
and the relatively weak correlation with respect to position.

To evaluate the effect of the systematic uncertainties on the reconstructed parameters, the probability
distributions for each signal and background were reconstructed with each parameter varied by
one standard deviation in each direction as given in Table \ref{tab:systematics}. The full data was
then refitted with each of the new sets of PDFs, and the difference between the new best-fit signal
and the nominal best-fit taken as the systematic uncertainty on the number of signal counts. The
nominal likelihood function was then convolved with an asymmetric normal distribution, whose width
was determined by the quadrature sum of these systematic uncertainties. The signal upper limit is
then determined by directly integrating the likelihood function, with a positive uniform prior in rate, 
up to the 90\% interval.

\begin{table}[ht]
    \centering
    \begin{tabular}{c|c|c|c|c|c}
        \hline \hline
        Data & \multicolumn{5}{c}{Signal Efficiency (\%)} \\
        Set & $n$ & $p$ & $pp$ & $np$ & $nn$ \\
        \hline
        1 & $11.4^{+0.7}_{-0.7}$ & $13.2^{+0.7}_{-0.7}$ & $11.5^{+0.5}_{-0.5}$ & $6.6^{+0.3}_{-0.3}$ & $1.84^{+0.07}_{-0.06}$\\
        2 & $11.6^{+0.7}_{-0.7}$ & $13.3^{+0.7}_{-0.7}$ & $11.5^{+0.5}_{-0.5}$ & $6.6^{+0.3}_{-0.3}$ & $1.84^{+0.07}_{-0.06}$\\
        3 & $11.5^{+0.7}_{-0.7}$ & $13.3^{+0.7}_{-0.7}$ & $11.5^{+0.5}_{-0.5}$ & $6.6^{+0.3}_{-0.3}$ & $1.84^{+0.07}_{-0.06}$\\
        4 & $10.9^{+0.7}_{-0.6}$ & $12.6^{+0.6}_{-0.6}$ & $10.8^{+0.5}_{-0.5}$ & $6.2^{+0.3}_{-0.3}$ & $1.72^{+0.07}_{-0.05}$\\
        5 & $14.6^{+0.9}_{-0.8}$ & $16.8^{+0.8}_{-0.8}$ & $14.4^{+0.6}_{-0.6}$ & $8.3^{+0.4}_{-0.4}$ & $2.31^{+0.09}_{-0.07}$\\
        6 & $13.9^{+0.4}_{-0.4}$ & $16.4^{+0.4}_{-0.4}$ & $14.2^{+0.3}_{-0.3}$ & $8.2^{+0.2}_{-0.2}$ & $2.38^{+0.04}_{-0.02}$\\
        \hline \hline
    \end{tabular}
    \caption{The acceptance efficiency of the signal after applying analysis cuts and the individual nuclear branching ratios. The reconstruction systematic uncertainties are given alongside each efficiency.}
    \label{tab:efficiency}
\end{table}

\section{Results}
Results for the upper limit on the five decay modes of interest, including systematic
uncertainties, are evaluated independently and given in Table \ref{tab:resultsTable}.
The total $\mathrm{T}_e$ and $\cos{\theta_\mathrm{sun}}$ spectra for the combined dataset for the neutron decay
mode are shown in Fig. \ref{fig:energyfit}, where the bands on the total spectrum show the effect
that the systematic uncertainties have on the final spectral shape. The signal likelihood
is shown in Fig. \ref{fig:likelihood}, which in each of the studied decay modes is consistent
with no observed signal.

\begin{table}
    \centering
    \begin{tabular}{c|c|c}
        \hline \hline
        Decay Mode & Partial Lifetime Limit & Existing Limits\\
        \hline
        $n$  & $9.0\times10^{29}$ y & $5.8\times10^{29}$ y \cite{KamLANDNucleonDecay}\\
        $p$  & $9.6\times10^{29}$ y & $3.6\times10^{29}$ y \cite{FirstNDPaper}\\
        $pp$ & $1.1\times10^{29}$ y & $4.7\times10^{28}$ y \cite{FirstNDPaper}\\
        $np$ & $6.0\times10^{28}$ y & $2.6\times10^{28}$ y \cite{FirstNDPaper}\\
        $nn$ & $1.5\times10^{28}$ y & $1.4\times10^{30}$ y \cite{KamLANDNucleonDecay}\\
        \hline \hline
    \end{tabular}
    \caption{Lifetime limits at 90\% C.I. for the invisible decay modes of interest alongside the existing limits.}
    \label{tab:resultsTable}
\end{table}


\section{Conclusion}
\label{sec:Conclusion}
Multiple datasets totaling 274.7 days of data-taking were analyzed to search for
five selected modes of ``invisible'' nucleon decay ($n$, $p$, $nn$, $pp$, $np$). 
Results are consistent with no observation of invisible nucleon decay
for all the studied decay modes, and represent an improvement on previous limits for all
but the $nn$ decay mode. In particular, the addition of the longer low-background data set
improved the previous \snoplus{} results on the $n$ and $p$ decay modes by a factor of 3 while 
improving the $np$ and $pp$ decay modes by a factor of 2 \cite{FirstNDPaper}. The much smaller
improvement found in the $nn$ decay mode was investigated and found to be due to it's energy
spectrum. Compared with the other four modes, the cascade of gamma-rays from $nn$ extends out
to higher energies, making the signal less distinguishable from $^{8}$B solar neutrino events.
As can be seen in Fig. \ref{fig:energyfit}, the large number of events at high energy, and
the slight deficit of events in the solar direction lowers the sensitivity to the $nn$ decay
mode.
\begin{acknowledgements}
Capital construction funds for the SNO\raisebox{0.5ex}{\tiny\textbf{+}} experiment were provided by the Canada Foundation for Innovation (CFI) and matching partners. 
This research was supported by: 
{\bf Canada: }
Natural Sciences and Engineering Research Council, 
the Canadian Institute for Advanced Research (CIFAR), 
Queen's University at Kingston, 
Ontario Ministry of Research, Innovation and Science, 
 Alberta Science and Research Investments Program, 
Federal Economic Development Initiative for Northern Ontario,
Ontario Early Researcher Awards;
{\bf U.S.: }
Department of Energy Office of Nuclear Physics, 
National Science Foundation, 
Department of Energy National Nuclear Security Administration through the Nuclear Science and Security Consortium; 
{\bf UK: }
Science and Technology Facilities Council (STFC),
the European Union's Seventh Framework Programme under the European Research Council (ERC) grant agreement,
the Marie Curie grant agreement;
{\bf Portugal: }
Funda\c{c}\~{a}o para a Ci\^{e}ncia e a Tecnologia (FCT-Portugal);
{\bf Germany: }
the Deutsche Forschungsgemeinschaft;
{\bf Mexico: }
DGAPA-UNAM and Consejo Nacional de Ciencia y Tecnolog\'{i}a.

We thank the SNO\raisebox{0.5ex}{\tiny\textbf{+}} technical staff for their strong contributions.  We would like to thank SNOLAB and its staff for support through underground space, logistical and technical services. SNOLAB operations are supported by CFI and the Province of Ontario Ministry of Research and Innovation, with underground access provided by Vale at the Creighton mine site.

This research was enabled in part by support provided by WestGRID (www.westgrid.ca) and Compute Canada (www.computecanada.ca) in particular computer systems and support from the University of Alberta (www.ualberta.ca) and from Simon Fraser University (www.sfu.ca) and by the GridPP Collaboration, in particular computer systems and support from Rutherford Appleton Laboratory~\cite{gridpp, gridpp2}. Additional high-performance computing was provided through the ``Illume'' cluster funded by CFI and Alberta Economic Development and Trade (EDT) and operated by ComputeCanada and the Savio computational cluster resource provided by the Berkeley Research Computing program at the University of California, Berkeley (supported by the UC Berkeley Chancellor, Vice Chancellor for Research, and Chief Information Officer). Additional long-term storage was provided by the Fermilab Scientific Computing Division. Fermilab is managed by Fermi Research Alliance, LLC (FRA) under Contract with the U.S. Department of Energy, Office of Science, Office of High Energy Physics.

For the purposes of open access, the authors have applied a Creative Commons Attribution licence to any Author Accepted Manuscript version arising. Representations of the data relevant to the conclusions drawn here are provided within this paper.
\end{acknowledgements}
\bibliography{References}

\end{document}